\documentclass[english,conference, twocolumn]{IEEEtran}
\usepackage[T1]{fontenc}
\usepackage[latin9]{inputenc}
\usepackage{color}
\usepackage{float}
\usepackage{amsthm}
\usepackage{amsmath}
\usepackage{amssymb}
\usepackage{graphicx}

\makeatletter
\theoremstyle{definition}
\newtheorem{assumption}{Assumption}
  \theoremstyle{plain}
  \newtheorem{lem}{\protect\lemmaname}
  \theoremstyle{definition}
  \newtheorem{defn}{\protect\definitionname}
  \theoremstyle{plain}
  \newtheorem{thm}{\protect\theoremname}
\theoremstyle{definition}
\newtheorem{property}{Property}

\usepackage{cite}
\IEEEoverridecommandlockouts

\makeatother

\usepackage{babel}
\providecommand{\definitionname}{Definition}
\providecommand{\lemmaname}{Lemma}
\providecommand{\theoremname}{Theorem}

\begin{document}

\title{Decentralized formation control with connectivity maintenance and
collision avoidance under limited and intermittent sensing%
\thanks{$^{1}$Department of Mechanical and Aerospace Engineering, University
of Florida, Gainesville FL 32611-6250, USA Email:\{tenghu, kanzhen0322,
joelar, wdixon\}@ufl.edu%
}%
\thanks{$^{2}$Department of Electrical and Computer Engineering, University
of Florida, Gainesville FL 32611-6250, USA Email: wdixon@ufl.edu%
}%
\thanks{This research is supported in part by NSF award numbers 0901491, 1161260,
1217908, ONR grant number N00014-13-1-0151, and a contract with the
AFRL Mathematical Modeling and Optimization Institute. Any opinions,
findings and conclusions or recommendations expressed in this material
are those of the authors and do not necessarily reflect the views
of the sponsoring agency.%
}}

\author{Teng-Hu Cheng$^{1}$, Zhen Kan$^{1}$, Joel A. Rosenfeld$^{1}$,
and Warren E. Dixon$^{1,2}$ }
\maketitle
\begin{abstract}
A decentralized switched controller is developed for dynamic agents
to perform global formation configuration convergence while maintaining
network connectivity and avoiding collision within agents and between
stationary obstacles, using only local feedback under limited and
intermittent sensing. Due to the intermittent sensing, constant position
feedback may not be available for agents all the time. Intermittent
sensing can also lead to a disconnected network or collisions between
agents. Using a navigation function framework, a decentralized switched
controller is developed to navigate the agents to the desired positions
while ensuring network maintenance and collision avoidance.
\end{abstract}

\section{Introduction}

Multi-agent systems have found applications in a wide range of situations.
These include problems of consensus \cite{Ren2007a,Olfati-Saber2004,Olfati-Saber2007},
rendezvous\cite{Lin2007,Lin2007a,Hui2011}, and formation and flocking
of multiple agents\cite{Dong2008,Jadba2003,Lin2004,Fax2004}. In these
applications, a decentralized control structure has advantages over
a centralized structure including: computational efficiency, robustness,
and flexibility. Various decentralized approaches ( cf. \cite{Gennaro2006,Tanner2007,Tanner2005,Gouvea.Lizarralde.ea2013,Tanner.Jadbabaie.ea2003})
have been developed to perform cooperative objectives for a multi-agent
system; however, network connectivity problems are often neglected.
The loss of connectivity can arise through limited communications
and limited sensing ranges and angles, and it can result in collisions
as well as the loss of the formation or individual agents. 

The broad applicability of dynamic network topologies helps explain
a recent increase in its popularity. In particular the issues surrounding
network connectivity has been gaining more focus. In \cite{Meng2005}
and \cite{Ji.Egerstedt2007}, decentralized schemes addressing connectivity
issues for dynamic topologies of formation and rendezvous problems
were approached with a graph theoretic method. In these papers, the
authors used a convergence analysis based on LaSalle's invariant theorem
with common Lyapunov functions. In \cite{dimarogonas2008}, the network
connectivity issue was handled with a navigation function based controller
using bounded control inputs for a formation problem using both static
and dynamic graphs. However, these applications didn't consider the
problem of collision avoidance. Both network connectivity and collision
avoidance were addressed in \textcolor{black}{\cite{Kan.Dani.ea2012}},
but only a fixed network topology is considered.

In some formation control problems, communication is not necessary,
but in these cases local feedback information from sensors is required.
Moreover, due to environment factors or limitations in the field-of-view
of sensors, the interaction graphs can be intermittent and time-varying.
Intermittent sensing problems were considered for formation control
problems using graph-theoretic methods in \cite{Lin.Francis.ea2005}
and \cite{Qu2009}. These problems were solved based on the existence
of a globally reachable node, but they didn't account for connectivity
or collision avoidance problems. A switched control scheme is developed
in \cite{Lin2004} for formation problems, but the controller neglects
network connectivity. In \cite{Cortes2009}, a coordination algorithm
was designed to stabilize the shape of the formation in a way that
it was robust to the sensing link failure, but connectivity and collision
avoidance were not considered. In \cite{dimarogonas2008_R} swarm
aggregation problems were investigated within fixed and dynamic network
topologies for both network connectivity and collision avoidance,
but the dynamic topologies only resulted from link additions to the
network. In \cite{Lin2007} and \cite{Xiao.Wang.ea2012}, decentralized
controllers were designed to address network connectivity. Unfortunately,
the control strategies were specific to rendezvous problems, and collision
avoidance was not considered. 

This paper considers formation control problems under limited and
intermittent sensing. Based on a navigation function framework, a
decentralized hybrid controller is developed to ensure network connectivity
and collision avoidance while controlling the formation. Nonsmooth
navigation functions are used which result in the use of a common
Lyapunov function, so the formation error of the entire configuration
converges globally with sufficiently small error (i.e. converges to
the neighborhood of the critical points) under arbitrary switching
sequence. This paper is organized as follows. In Section \ref{sec:Problem-Formulation},
the dynamics of the agents and the problem are formulated. Then the
navigation function based controller is proposed in Section \ref{sec:Control-development}.
We perform a connectivity analysis in Section \ref{sec:Connectivity-Analysis}
and a convergence analysis in Section \ref{sec:Convergence-Analysis}.
Finally, the simulation results are presented in Section \ref{sec:Simulation}.

\section{\label{sec:Problem-Formulation}Problem Formulation}

Consider $N$ dynamic point-mass agents in the finite workspace $\mathcal{F}\subset\mathbb{R}^{2}$
with motion governed by the following kinematics
\begin{equation}
\dot{q}_{i}=u_{i},\; i=1,\ldots,N\label{eq:dynamic}
\end{equation}
where $q_{i}\in\mathbb{R}^{2}$ represents the position of agent $i$
in a two-dimensional space, and $u_{i}\in\mathbb{R}^{2}$ denotes
the control input of agent $i$. The subsequent development is based
on the assumption that each agent has a limited sensing range, which
is encoded by a disk centered at the agent. Position feedback is only
available for agents within the interior of the disk. Moreover, the
sensing is assumed to be intermittent (i.e., existing links within
the disk region may fail), which implies that two agents do not have
continuous state feedback even if they remain within the sensing zone
of each other. 

Since sensing is intermittent, the set of neighbor nodes that can
be successfully sensed by agent $i$ at $t\in\mathbb{R}_{\geq0}$
is denoted as the time-varying set $\mathcal{N}_{i}^{s}(t)$, where
$\mathcal{N}_{i}^{s}:\left[0,\,\infty\right)\rightarrow\mathcal{V},$
where $\mathcal{V}\triangleq\left\{ 1,\,2,\cdots,\, N\right\} $ is
an index set of all agents in the system. As a result, the sensor
graph of the network system is an undirected, time-varying graph that
can be modeled as $\mathcal{G}(t)=\left(\mathcal{\mathcal{V}},\,\mathcal{E}(t)\right)$,
where $\mathcal{E}(t)\triangleq\left\{ \left(i,\, j\right)\in\mathcal{V}\times\mathcal{V}|\, j\in\mathcal{N}_{i}^{s}(t),\, i\in\mathcal{V},\, i\neq j\right\} $,
where $i$ and $j$ represent nodes located at $q_{i}$ and $q_{j}$,
$d_{ij}\in\mathbb{R}_{\geq0}$ denotes the distance between two nodes
defined as $d_{ij}\triangleq\left\Vert q_{i}-q_{j}\right\Vert ,$
and $R_{s}$ is the maximal sensing radius for every agent. To include
all the time-varying graphs, a switched graph is defind as $G_{\sigma(t)},$
where $\sigma:\left[0,\,\infty\right)\rightarrow\mathcal{P}$ is a
switching signal, and $\mathcal{P}\in\left\{ 1,\,2,\,\ldots,\, P\right\} $
is a finite index set such that $\left\{ G_{p}:\, p\in\mathcal{P}\right\} $
includes all possible graphs $\underset{t\geq0}{\cup}\mathcal{G}(t)$. 

Network connectivity maintenance is ensured by preserving every existing
link in the network. Particularly, the agents are considered connected
if they stay within the sensing zone of the desired neighboring agents
(even if there are intermittent sensing link failures), if they are
neighbors initially, i.e.,
\begin{equation}
d_{ij}\left(t\right)<R_{s},\:\forall t\geq0.\label{eq:connectivity}
\end{equation}
The objective in this paper is to maintain network connectivity while
also achieving a desired formation, which is specified by 
\begin{equation}
\left\Vert q_{i}-q_{j}-c_{ij}\right\Vert \rightarrow0\;\mbox{as}\; t\rightarrow\infty,\; j\in\mathcal{N}_{i}^{f},\, i\in\mathcal{V},\label{eq:formation formula}
\end{equation}
where $\mathcal{N}_{i}^{f}$ is the set of preassigned agents, and
$c_{ij}\in\mathbb{R}^{2},$ satisfying $c_{ij}=-c_{ji},$ describes
the desired relative position between node $i$ and the adjacent node
$j\in\mathcal{N}_{i}^{f}$. Different from $\mathcal{N}_{i}^{s}(t)$
which is time-varying due to the intermittent sensing, $\mathcal{N}_{i}^{f}$
is time-invariant.

Consider stationary obstacles $o_{1},\, o{}_{2},\ldots,\, o_{m}$
in the workspace $\mathcal{F}$, which are represented by a set of
$m$ points indexed by $\mathcal{M}=\left\{ 1,\,2,\cdots,\, m\right\} $.
To prevent collisions among agents and obstacles, a disk region centered
at agent $i$ with radius $\delta_{1}<R_{s}$ is defined. Any agent
or obstacle in this region is considered as a potential collision
with agent $i$, and the potential collision set $\mathcal{N}_{i}:\left[0,\,\infty\right)\rightarrow\mathcal{V}$
is defined as
\begin{equation}
\mathcal{N}_{i}\left(t\right)\triangleq\left\{ j\in\mathcal{V}\mid\left\Vert q_{i}-q_{j}\right\Vert \leq\delta_{1}\right\} .\label{eq:obstacle set}
\end{equation}

Since only $j\in\mathcal{N}_{i}^{f}$ in (\ref{eq:connectivity})
are required to maintain the neighborhood with agent $i$ in the desired
formation, (\ref{eq:connectivity}) can be modified as
\begin{equation}
d_{ij}\left(t\right)<R_{s},\:\forall t\geq0,\; j\in\mathcal{N}_{i}^{f},\, i\in\mathcal{V}.\label{eq:modified connectivity}
\end{equation}

In summary, the objective is to asymptotically achieve a formation
configuration as in (\ref{eq:formation formula}), while ensuring
network connectivity as in (\ref{eq:modified connectivity}) and collision
avoidance between agents and stationary obstacles $o_{1},\, o{}_{2},\ldots,\, o_{m}$.

\begin{assumption}
\label{finite switchings}The sensing link failures between agents
happen a finite number of times in a finite time interval, (i.e.,
the switching signal $\sigma$ has finite switches in any finite time
interval.) Specifically, given any non-overlapping time interval $\left[t_{k},\, t_{k+1}\right)$,
$k=0,\,1,\,\cdots$, then $0<\tau<t_{k+1}-t_{k}<T,$ where $\tau\in\mathbb{R}$
is the non-vanishing dwell-time, and $T\in\mathbb{R}$ is a positive
constant. The graph $G_{\sigma}$ is invariant for $t\in\left[t_{k},\, t_{k+1}\right),$
$\forall\, k=0,\,1,\,2,\,\cdots$, and the switching sequence of $\sigma$
is arbitrary. 
\end{assumption}

\begin{assumption}
\label{assumption1}The desired formation neighbor set of agent $i$
is initially inside its sensing zone, $\mathcal{N}_{i}^{f}\subset\mathcal{N}_{i}^{s}(t_{0}),$
$\forall i\in\mathcal{V},$ and the neighboring agents are not initially
located at any unstable equilibria.
\end{assumption}

\begin{assumption}
\label{aummption2}The desired relative position described by $c_{ij}$
is achievable (i.e., $\delta_{1}<\left\Vert c_{ij}\right\Vert <R_{s}-\delta_{2}$,
where $\delta_{2}\in\mathbb{R}^{+}$ denotes a buffer distance for
connectivity maintenance. So the relative position would not result
in a partition of the graph or cause collision of any two agents.)
and the agents do not take certain pathological configurations. One
example would be all of the agents and goals being co-linear. However
this and other such configurations are assumed to constitute a Lebesgue
measure zero set in the space of all configurations, and are practically
resolved by small perturbations.
\end{assumption}

\section{\label{sec:Control-development}Control development}

Based on \cite{Kan.Dani.ea2012}, a navigation function $\varphi_{i}:\mathcal{F}\rightarrow\left[0,\,1\right]$
for each agent $i$ is designed as,
\begin{equation}
\varphi_{i}=\frac{\gamma_{i}}{\left(\gamma_{i}^{k}+\beta_{i}\right)^{\frac{1}{k}}},\label{eq:potential function}
\end{equation}
where $k\in\mathbb{R}$ is an adjustable positive constant, $\gamma_{i}:\mathbb{R}^{2}\rightarrow\mathbb{R}_{\geq0}$
is a goal function, and $\beta_{i}:\mathbb{R}_{\geq0}\rightarrow\left[0,\,1\right]$
is a constraint function for agent $i$. Based on the objective in
(\ref{eq:formation formula}), the goal function $\gamma_{i}$ in
(\ref{eq:potential function}) is designed as
\begin{equation}
\gamma_{i}(q_{i},\, q_{j})\triangleq\underset{j\in\mathcal{N}_{i}^{f}}{\sum}\left\Vert q_{i}-q_{j}-c_{ij}\right\Vert ^{2}.\label{eq:gamma i}
\end{equation}
The constraint function $\beta_{i}$ is defined as
\begin{equation}
\beta_{i}\triangleq\underset{j\in\mathcal{N}_{i}^{f}}{\prod}b_{ij}\underset{k\in\mathcal{N}_{i}\cup\mathcal{M}_{i}}{\prod}B_{ik},\label{eq:constraint func}
\end{equation}
which enables collision avoidance and connectivity maintenace. To
maintain network connectivity, the nonsmooth function $b_{ij}:\mathbb{R}_{\geq0}\rightarrow[0,\,1]$
in (\ref{eq:constraint func}) is designed as
\begin{equation}
b_{ij}\left(d_{ij}\right)\triangleq\left\{ \begin{array}{cc}
1, & d_{ij}<R_{s}-\delta_{2},\\
\begin{array}{c}
-\frac{1}{\delta_{2}^{2}}(d_{ij}+2\delta_{2}-R_{s})^{2}\\
+\frac{2}{\delta_{2}}(d_{ij}+2\delta_{2}-R_{s}),
\end{array} & \begin{array}{c}
R_{s}-\delta_{2}\leq d_{ij}\leq R_{s},\end{array}\\
0, & d_{ij}>R_{s},
\end{array}\right.\label{eq:bij}
\end{equation}
where $b_{ij}$ is not differentiable at $R_{s}.$ Specifically, $b_{ij}$
is designed to prevent node $i$ from leaving the communication region
of its formation neighbor $j\in\mathcal{N}_{i}^{f}$. Let $\mathcal{M}_{i}$
denote the set of stationary obstacles within the collision region
of node $i$. In (\ref{eq:constraint func}), for each node $k\in\mathcal{N}_{i}\cup\mathcal{M}_{i}$,
$B_{ik}:\mathbb{R}\rightarrow\left[0,\,1\right]$ is defined as 
\[
B_{ik}\left(d_{ik}\right)\triangleq\left\{ \begin{array}{cc}
-\frac{1}{\delta_{1}^{2}}d_{ik}^{2}+\frac{2}{\delta_{1}}d_{ik}, & \;0\leq d_{ik}\leq\delta_{1},\\
1, & \; d_{ik}>\delta_{1}.
\end{array}\right.
\]
Therefore, $\beta_{i}\rightarrow0$ when node $i$ enters the constraint
region, (i.e. when node $i$ approaches other nodes, stationary obstacles,
or tries to leave the sensing range of their adjacent nodes $j\in\mathcal{N}_{i}^{f},\:\forall t\geq0$). 

Based on Assumption \ref{aummption2}, $\gamma_{i}$ and $\beta_{i}$
will not be zero at the same time, and the navigation function $\varphi_{i}$
reaches its maximum at $1$ when $\beta_{i}=0$ and its minimum at
$0$ when $\gamma_{i}=0$. 

Due to the intermittent sensing, consider the two sets $\mathcal{V}_{f}(t)$
and $\mathcal{V}_{u}(t)$, where $\mathcal{V}_{f},\,\mathcal{V}_{u}:\left[0,\,\infty\right)\rightarrow\mathcal{V}$
are defined as $\mathcal{V}_{f}(t)\triangleq\left\{ i\in\mathcal{V}|\,\mathcal{N}_{i}^{f}=\mathcal{N}_{i}^{s}(t)\cap\mathcal{N}_{i}^{f}\right\} $
and $\mathcal{V}_{u}(t)\triangleq\mathcal{V}\setminus\mathcal{V}_{f}(t).$
The set $\mathcal{V}_{f}(t)$ includes agents that can sense all of
the formation neighbors $\mathcal{N}_{i}^{f}$ at $t\in\mathbb{R}_{\geq0}.$
Otherwise, agent $i$ will be in $\mathcal{V}_{u}(t)$ at some $t\in\mathbb{R}_{\geq0}$.
Using the navigation function in (\ref{eq:potential function}), the
decentralized switched controller for agent $i$ is designed as
\begin{equation}
u_{i}(t)=\left\{ \begin{array}{cc}
-\Gamma\nabla_{q_{i}}\varphi_{i},\quad & i\in\mathcal{V}_{f}(t),\\
0, & i\in\mathcal{V}_{u}(t),
\end{array}\right.\label{eq:controller}
\end{equation}
where $\Gamma\in\mathbb{R}^{+}$ is a positive constant gain, and
$\nabla_{q_{i}}\left(\cdot\right)\triangleq\frac{\partial}{\partial q_{i}}\left(\cdot\right).$
In (\ref{eq:controller}), the control switching scheme of agent $i$
is based on the sensing condition at time $t$. If all neighbor agents
in $\mathcal{N}_{i}^{f}$ can be sensed by agent $i$, $u_{i}(t)=-\Gamma\nabla_{q_{i}}\varphi_{i}$,
and $u_{i}(t)=0$ otherwise.

\section{\label{sec:Connectivity-Analysis}Connectivity Analysis}
\begin{lem}
If the initial graph of the multi-agent system is connected, then
the controller in ($\ref{eq:controller}$) ensures agent $i$ and
$j$ remain connected for all time.\end{lem}
\begin{IEEEproof}
Consider an agent $i\in\mathcal{V}$ located at $q_{0}\in\mathcal{F},$
where the sensing link is about to break, which implies $\underset{j\in\mathcal{N}_{i}^{f}}{\prod}b_{ij}\rightarrow0$,
then three cases must be considered.

\textit{Case 1.} As agent $j\in\mathcal{N}_{i}^{f}$ approaches the
sensing region (i.e., $\left\Vert q_{i}-q_{j}\right\Vert $ approaches
$R_{s}$ from the left), then $\beta_{i}$ tends to zero. The gradient
of $\varphi_{i}$ is 
\begin{equation}
\nabla_{q_{i}}\varphi_{i}=\frac{k\beta_{i}\nabla_{q_{i}}\gamma_{i}-\gamma_{i}\nabla_{q_{i}}\beta_{i}}{k(\gamma_{i}^{k}+\beta_{i})^{\frac{1}{k}+1}}.\label{eq:grad potential}
\end{equation}
Consider
\begin{align*}
\nabla_{q_{i}}\beta_{i}= & \underset{h\in\mathcal{N}_{i}^{f}}{\sum}\underset{l\neq h}{\underset{l\in\mathcal{N}_{i}^{f},}{\Pi}}b_{il}\left(\nabla_{q_{i}}b_{ih}\right)\underset{k\in\mathcal{N}_{i}\cup\mathcal{M}_{i}}{\prod}B_{ik}\\
 & +\underset{h\in\mathcal{N}_{i}\cup\mathcal{M}_{i}}{\sum}\underset{j\in\mathcal{N}_{i}^{f}}{\prod}b_{ij}\underset{l\neq h}{\underset{l\in\mathcal{N}_{i}\cup\mathcal{M}_{i},}{\Pi}}B_{il}\left(\nabla_{q_{i}}B_{ih}\right).
\end{align*}
Provided only agent $j$ is near the boundary (i.e., $\left\Vert q_{i}-q_{j}\right\Vert \rightarrow R_{s}^{-}$),
$\nabla_{q_{i}}\beta_{i}$ has only one dominant term:
\begin{align*}
\nabla_{q_{i}}\beta_{i}= & \underset{l\neq j}{\underset{l\in\mathcal{N}_{i}^{f},}{\Pi}}b_{il}\left(\nabla_{q_{i}}b_{ij}\right)\underset{k\in\mathcal{N}_{i}\cup\mathcal{M}_{i}}{\prod}B_{ik}+O\left(b_{ij}\right),
\end{align*}
where $O\left(\cdot\right)$ is the Big O notation, which vanishes
as $b_{ij}$ approaches $R_{s}.$ The other term in the numerator
of $\nabla_{q_{i}}\varphi_{i}$ in (\ref{eq:grad potential}) is $k\beta_{i}\nabla_{q_{i}}\gamma_{i}=O\left(b_{ij}\right),$
hence $\nabla_{q_{i}}\varphi_{i}$ in (\ref{eq:grad potential}) can
be expressed as 
\begin{align*}
\nabla_{q_{i}}\varphi_{i} & =\\
 & \frac{-\gamma_{i}\underset{l\neq j}{\underset{l\in\mathcal{N}_{i}^{f},}{\Pi}}b_{il}\underset{k\in\mathcal{N}_{i}\cup\mathcal{M}_{i}}{\prod}B_{ik}\left(\nabla_{q_{i}}b_{ij}\right)+O\left(b_{ij}\right)}{k(\gamma_{i}^{k}+\beta_{i})^{\frac{1}{k}+1}}.
\end{align*}
Note that the gradient of $b_{ij}$ w.r.t. $q_{i}$ can be determined
as 
\begin{equation}
\nabla_{q_{i}}b_{ij}=\left\{ \begin{array}{cc}
0, & \begin{array}{c}
d_{ij}<R_{s}-\delta_{2}\:\mathrm{or}\\
d_{ij}>R_{s},
\end{array}\\
-\frac{2(d_{ij}+\delta_{2}-R_{s})\left(q_{i}-q_{j}\right)}{\delta_{2}^{2}d_{ij}}, & R_{s}-\delta_{2}\leq d_{ij}<R_{s},
\end{array}\right.\label{eq:gradient b_ij}
\end{equation}
where $\gamma_{i}$, $b_{il}$, $B_{ik},$ $k,$ $\delta_{2},$ and
$R_{s}$ are positive constants. Thus, $\dot{q}_{i}=-\Gamma\nabla_{q_{i}}\varphi_{i}$
points in the direction of $q_{j}-q_{i},$ which forces nodes $i$
to move toward node $j$.

\textit{Case 2.} Now suppose several agents $j_{1},\, j_{2},\,\ldots,\, j_{s}\in\mathcal{N}_{i}^{f}$
are near the boundary of the sensing region. That is, $d_{ij_{m}}$
is near $R_{s}$ for each $m=1,\,2,\,\ldots,\, s.$ For this case,
$\nabla_{q_{i}}\varphi_{i}=\frac{-\gamma_{i}\underset{m}{\sum}\underset{l\neq j_{m}}{\underset{l\in\mathcal{N}_{i}^{f},}{\prod}}b_{il}\underset{k\in\mathcal{N}_{i}\cup\mathcal{M}_{i}}{\prod}B_{ik}\left(\nabla_{q_{i}}b_{ij_{m}}\right)}{k(\gamma_{i}^{k}+\beta_{i})^{\frac{1}{k}+1}}+O\left(\underset{m}{\prod}b_{ij_{m}}\right).$
The first term above in $\nabla_{q_{i}}\varphi_{i}$ tends to zero,
however since the $b_{ij_{m}}$ terms are quadratic near $R_{s}$,
the order of the zero contributed by the first term is one degree
less than $O\left(\underset{m}{\prod}b_{ij_{m}}\right)$, so the first
term dominantes as each $d_{ij_{m}}\rightarrow R_{s}.$ Hence $\dot{q}_{i}=-\Gamma\nabla_{q_{i}}\varphi_{i}$
is approximately a linear combination of the vectors $q_{j_{1}}-q_{i},$
$q_{j_{2}}-q_{i},\ldots,\, q_{j_{s}}-q_{i},$ where the largest contribution
comes from those $j_{m}$ closest to the sensing boundary. Thus, node
$i$ moves almost toward $j_{m}$ resulting in a largest decrease
in $d_{ij_{m}},$ so the connectivity can be maintained. 

\textit{Case 3.} Consider a node $i\in\mathcal{V}_{u}$ (or more than
one node in the set of $\mathcal{V}_{u}$). The controller will be
$u_{i}=0$ based on (\ref{eq:controller}). Since both node $i$ and
its neighbor $j\in\mathcal{N}_{i}^{f}$ are in the undirected graph,
node $j$ can't sense node $i$, so $j\in\mathcal{V}_{u},$ thus $u_{j}=0$.
Since both $i,\, j$ nodes have no control input, the distance between
them remains the same. 

By Assumption \ref{assumption1}, $\mathcal{N}_{i}^{f}\subset\mathcal{N}_{i}^{s}(t_{0}),\, i\in\mathcal{V}.$
Furthermore, from \textit{Case 1}-\textit{Case 3}, the decentralized
switched control policy in (\ref{eq:controller}) ensures the distances
between agent $i\in\mathcal{V}$ and its formation neighbors $j\in\mathcal{N}_{i}^{f}$
never increase under intermittent sensing conditions. As a result,
the formation neighbors $j\in\mathcal{N}_{i}^{f}$ remain inside the
sensing region of agent $i$ for all time. Specifically, 
\begin{equation}
d_{ij}\left(t\right)<R_{s},\, j\in\mathcal{N}_{i}^{f},\, i\in\mathcal{V},\,\forall t\geq0.\label{eq:connectivity eq}
\end{equation}

\end{IEEEproof}

\section{\label{sec:Convergence-Analysis}Convergence Analysis}
\begin{defn}
\label{def-K[]}\cite{Paden1987} Consider the following differential
equation with a discontinuous right-hand side:
\begin{equation}
\dot{x}=f(x),\label{eq:differential equation}
\end{equation}
where $f:\mathbb{R}^{n}\rightarrow\mathbb{R}^{n}$ is measurable and
essentially locally bounded, and $n\in\mathbb{N}$ is a finite constant.
The vector function $x$ is called a solution of $\left(\ref{eq:differential equation}\right)$
on $\left[t_{0},\, t_{1}\right]$ if $x$ is absolutely continuous
on $\left[t_{0},\, t_{1}\right]$ and for almost all $t\in\left[t_{0},\, t_{1}\right]$
\[
\dot{x}\in K\left[f\right]\left(x\right)
\]
\begin{equation}
K\left[f\right]\left(x\right)\triangleq\underset{\delta>0}{\cap}\underset{\mu N=0}{\cap}\overline{co}\, f\left(B\left(x,\,\delta\right)\setminus N\right),\label{Philippov def  K[]}
\end{equation}
where $\underset{\mu N=0}{\cap}$ denotes the intersection over all
sets $N$ of Lebesgue measure zero. 
\end{defn}

To prove the convergence of the agents to the desired formation, an
invariance principle for switched systems is applied to a common Lyapunov
function candidate $V:\mathbb{R}^{2N}\rightarrow\mathbb{R}$ given
by 
\begin{equation}
V(q)\triangleq\sum_{i=1}^{N}\varphi_{i},\label{eq:Lyapunov}
\end{equation}
 where $q$ is the stack state vector, and $V$ reaches its minimum
value of $0$ if the desired formation is achieved.

\begin{thm}
\label{[thm] V_ae}\cite{Shevitz1994} Let $x\left(\cdot\right)$
be a Filippov solution to $\dot{x}=f(x)$ on an interval containing
$t$ and $V:\mathbb{R}^{n}\rightarrow\mathbb{R}$ be a Lipschitz and
regular function. Then $V\left(x\left(t\right)\right)$ is absolutely
continuous, $\frac{d}{dt}V\left(x(t)\right)$ exists almost everywhere
(a.e.) and
\[
\frac{d}{dt}V\left(x\left(t\right)\right)\overset{a.e.}{\in}\dot{\tilde{V}}\left(x\right)\triangleq\underset{\xi\in\partial V\left(x\left(t\right)\right)}{\cap}\xi^{T}K\left[f\right]\left(x\left(t\right)\right).
\]

\end{thm}

Based on Definition \ref{def-K[]} and Theorem \ref{[thm] V_ae} ,
the main result of this paper is provided as follows.
\begin{thm}
Given (\ref{eq:controller}), the maximum relative position errors
of any two formation neighbors of the network system in (\ref{eq:dynamic})
converges to $\underset{j\in\mathcal{N}_{i}^{f}}{\max}\left\Vert q_{i}-q_{j}-c_{ij}\right\Vert =\sqrt{\frac{c_{\mbox{max}}}{\underline{N}}},\, i\in\mathcal{V}$
provided that the adjustable gain $k$ in (\ref{eq:potential function})
is selected sufficiently large and every agent can sense all its formation
neighbors in the finite time interval $\underset{t\in\left[t_{k},\, t_{k+n}\right)}{\cup}\left(\mathcal{N}_{i}^{f}\cup\left\{ i\right\} \right)=\mathbb{\mathcal{V}},$
where $n\in\mathbb{N}$ is finite. \end{thm}
\begin{IEEEproof}
Consider the common Lyapunov function candidate $V$ defined in (\ref{eq:Lyapunov}),
where $V$ can be minimized at the critical points as shown in \cite{Kan.Dani.ea2012},
and $V$ reaches its minimum value of $0$ when the desired formation
is achieved. Based on Theorem \ref{[thm] V_ae}, 
\begin{equation}
\frac{d}{dt}V\left(q\left(t\right)\right)\overset{a.e.}{\in}\dot{\tilde{V}}\left(q\right)\triangleq\underset{\xi\in\partial V\left(x\left(t\right)\right)}{\cap}\xi^{T}K\left[\dot{q}\right].\label{eq:V_dot ae}
\end{equation}
The finite sums property of the generalized gradient defined in \cite{clarke1983}
gives 
\begin{equation}
\partial V\subset\left[\partial_{q_{1}}V^{T},\,\partial_{q_{2}}V^{T},\,\ldots,\,\partial_{q_{N}}V^{T}\right]^{T}.\label{eq:generalized gradient}
\end{equation}
Using (\ref{eq:V_dot ae}) and (\ref{eq:generalized gradient}), the
generalized time derivative of $V$ in (\ref{eq:V_dot ae}) can be
expressed as 
\begin{align}
\dot{\tilde{V}} & \subset\sum_{i\in\mathcal{V}}\left(\underset{\xi_{i}}{\cap}\,\xi_{i}^{T}\, K\left[\dot{q_{i}}\right]\right).\label{eq:V_dot0}
\end{align}
where $\xi_{i}\in\partial_{q_{i}}V.$ To turn the generalized gradient
into the gradient, the points at which $V$ is not differentiable
and Lebesgue measure zero need to be considered. From the inequality
in (\ref{eq:connectivity eq}), $d_{ij}$ never takes on the value
$d_{ij}=R_{s},\, j\in\mathcal{N}_{i}^{f},\, i\in\mathcal{V},$ at
the nonsmooth point of $b_{ij},$ so $b_{ij}$ is differentiable w.r.t.
$q_{i}$ along the solution of the closed-loop system. Since $B_{ik}$
and $\gamma_{i}$ are differentiable functions, $V$ is differentiable
w.r.t. $q_{i}$ along the solution of the closed-loop system for $i\in\mathcal{V}$.
Therefore, the generalized gradient can be expressed as 
\begin{equation}
\partial_{q_{i}}V\left(q\right)=\left\{ \nabla_{q_{i}}V\left(q\right)\right\} ,\, i\in\mathcal{V}.\label{eq:gradient_convert}
\end{equation}
Based on (\ref{eq:gradient_convert}), (\ref{eq:V_dot0}) can be rewritten
as 
\begin{equation}
\dot{\tilde{V}}\subset\sum_{i\in\mathcal{V}}\left(\nabla_{q_{i}}V{}^{T}K\left[\dot{q_{i}}\right]\right).\label{eq:V_dot1}
\end{equation}
By segregating $\mathcal{V}$ into the sets, $\mathcal{V}_{f}$ and
$\mathcal{V}_{u},$ (\ref{eq:V_dot1}) can be rewritten as
\begin{align}
\dot{\tilde{V}} & \subset\sum_{i\in\mathcal{V}_{f}}\left(\nabla_{q_{i}}V{}^{T}K\left[\dot{q_{i}}\right]\right)+\sum_{i\in\mathcal{V}_{u}}\left(\nabla_{q_{i}}V{}^{T}K\left[\dot{q_{i}}\right]\right).\label{eq:V_dot1-1}
\end{align}
From Assumption \ref{finite switchings}, the switching graph $G_{\sigma\left(t\right)}$
is invariant for $t\in\left[t_{k},\, t_{k+1}\right),$ so the set
$\mathcal{V}_{f}$ is also invariant during that time period. Based
on the switched control scheme in (\ref{eq:controller}), the second
term on the RHS of (\ref{eq:V_dot1-1}) will be zero, therefore,
\begin{equation}
\dot{\tilde{V}}\subset\sum_{i\in\mathcal{V}_{f}}\left(\nabla_{q_{i}}V{}^{T}K\left[\dot{q_{i}}\right]\right),\, t\in\left[t_{k},\, t_{k+1}\right).\label{eq:V_dot1-1a}
\end{equation}
In addition, by the definition of $K\left[\dot{q_{i}}\right]$ in
(\ref{Philippov def  K[]}), the switched controller in (\ref{eq:controller})
can be expressed as
\begin{align}
K\left[\dot{q_{i}}\right]\subset\, & \overline{co}\left\{ -\Gamma\nabla_{q_{i}}\varphi_{i}\,,\left[\begin{array}{c}
0\\
0
\end{array}\right]\right\} .\label{eq:K[]}
\end{align}
Also based on Assumption \ref{finite switchings}, the switching time
instance is Lebesgue measure zero, so (\ref{eq:K[]}) can be further
expressed as $K\left[\dot{q_{i}}\right]\subset\left\{ -\Gamma\nabla_{q_{i}}\varphi_{i}\right\} .$
Thus, by using the gradient of $V,$ (\ref{eq:V_dot ae}) and (\ref{eq:V_dot1-1a})
can be used to conclude that 
\begin{align}
\dot{V}\overset{a.e.}{\leq} & -\sum_{i\in\mathcal{V}_{f}}\left(\Gamma\left(\sum_{j=1}^{N}\nabla_{q_{i}}\varphi_{j}\right)^{T}\nabla_{q_{i}}\varphi_{i}\right),\label{eq:V_dot3}
\end{align}
where $t\in\left[t_{n},\, t_{n+1}\right),\, n\in\mathbb{N}.$ An equivalent
way to prove $\dot{V}\overset{a.e.}{<}0$ is to show $\sum_{i\in\mathcal{V}_{f}}\left(\Gamma\left(\sum_{j=1}^{N}\nabla_{q_{i}}\varphi_{j}\right)^{T}\nabla_{q_{i}}\varphi_{i}\right)>0,$
and based on the development in the appendix, its sufficient condition
is 
\begin{align}
 & \sum_{i\in\mathcal{V}_{f}}\left(4\underline{\beta}\left\Vert \underset{j\in\mathcal{N}_{i}^{f}}{\sum}\left(q_{i}-q_{j}-c_{ij}\right)\right\Vert ^{2}-\frac{\rho_{1,i}}{2k}-\frac{\rho_{2,i}}{2k^{2}}\right)>0,\label{eq:V_dot4-1}
\end{align}
for $t\in\left[t_{n},\, t_{n+1}\right).$ In (\ref{eq:V_dot4-1}),
$\rho_{1,i},$ $\rho_{2,i}\in\mathbb{R}$ are functions defined as
$\rho_{1,i}\triangleq c_{1,i}\gamma_{i}+c_{2,i}\gamma_{i}^{2}+c_{3,i}\left(\sum_{k=1}^{N}\gamma_{k}\right)^{2}$,
$\rho_{2,i}\triangleq c_{4,i}\gamma_{i}^{2}+c_{5,i}\left(\sum_{k=1}^{N}\gamma_{k}\right)^{2},$
where $c_{p,i}\in\mathbb{R},\, p=1-5,$ are positive constants. To
develop a further sufficient condition for (\ref{eq:V_dot4-1}), we
exploit the facts from \cite{Tanner.Kumar2005} that $\nabla_{q_{i}}\gamma_{i}\triangleq2\underset{j\in\mathcal{N}_{i}^{f}}{\sum}\left(q_{i}-q_{j}-c_{ij}\right)$
and $\left\Vert \nabla_{q_{i}}\gamma_{i}\right\Vert \geq\frac{\gamma_{i}}{R},$
where $R\triangleq\max\left\Vert q_{i}-q_{j}\right\Vert ,\, q_{i},\, q_{j}\in\mathcal{F},$
$\forall j\in\mathcal{N}_{i}^{f}.$ Hence, from (\ref{eq:connectivity eq})
\begin{equation}
\left\Vert \nabla_{q_{i}}\gamma_{i}\right\Vert \geq\frac{\gamma_{i}}{R_{s}},\label{eq:gradient_gamma bounds}
\end{equation}
and a sufficient condition for (\ref{eq:V_dot4-1}) can be developed
as 
\begin{equation}
\sum_{i\in\mathcal{V}_{f}}\left(\underline{\beta}\frac{\gamma_{i}^{2}}{R_{s}^{2}}-\frac{\rho_{1,i}}{2k}-\frac{\rho_{2,i}}{2k^{2}}\right)>0.\label{eq:V_dot5}
\end{equation}
By solving (\ref{eq:V_dot5}) for $\gamma_{i}$ and using (\ref{eq:gamma i}),
a further sufficient condition for (\ref{eq:V_dot4-1}) is 
\begin{equation}
\underset{j\in\mathcal{N}_{i}^{f}}{\sum}\left\Vert q_{i}-q_{j}-c_{ij}\right\Vert ^{2}>c_{\mbox{max}},\, i\in\mathcal{V}_{f},\label{eq:Gamma UUB}
\end{equation}
where $c_{\mbox{max}}\triangleq\sqrt{\frac{R_{s}^{2}}{\underline{\beta}}\left(\frac{\overline{\rho}_{1}}{2k}+\frac{\overline{\rho}_{2}}{2k^{2}}\right)},$
and $\overline{\rho}_{1},$ $\overline{\rho}_{2},$ $\underline{\beta}\in\mathbb{R}_{>0}$
are positive constants defined as $\overline{\rho}_{1}\triangleq\underset{i\in\mathcal{V}}{\max}\rho_{1,i},$
$\overline{\rho}_{2}\triangleq\underset{i\in\mathcal{V}}{\max}\rho_{2,i},$
and $\underline{\beta}\triangleq\underset{j\in\mathcal{N}_{i}^{f},\, i\in\mathcal{V}}{\min}\beta_{i}\beta_{j}.$
Additionally, $\beta_{i},\,\beta_{j}\neq0$ due to the fact that no
open set of initial solutions can be attracted to the maxima of $\varphi_{i}$
(i.e., $\beta_{i}=0$) along the negative gradient motion $-\frac{\partial\varphi_{i}}{\partial q_{i}}$
\cite{Karagoz.Bozma.ea2003}. Recall that $V$ in (\ref{eq:Lyapunov})
is a common Lyapunov function, so the switching signal $\sigma$ of
the time-varying graphs $G_{\sigma}$ can have arbitrary sequence
provided that (\ref{eq:Gamma UUB}) holds. Additionally, (\ref{eq:Gamma UUB})
can be extended to global (i.e., $i\in\mathcal{V}$) formation configuration
convergence if the switching signal $\sigma$ switches in the way
that satisfies the following condition
\begin{equation}
\underset{t\in\left[t_{k},\, t_{k+n}\right)}{\cup}\mathbb{\mathcal{V}}_{f}=\mathbb{\mathcal{V}},\, n\in\mathbb{N},\label{eq:graph condition}
\end{equation}
where $n$ is a finite positive constant. Based on (\ref{eq:Gamma UUB}),
and the ultimate maximum formation error for the entire switched system
can be expressed as 
\begin{equation}
\underset{j\in\mathcal{N}_{i}^{f}}{\max}\left\Vert q_{i}-q_{j}-c_{ij}\right\Vert =\sqrt{\frac{c_{\mbox{max}}}{\underline{N}}},\, i\in\mathcal{V},\label{eq:UUB distance}
\end{equation}
where $\underline{N}\triangleq\underset{i\in\mathcal{V}}{\min}\left|\mathcal{N}_{i}^{f}\right|.$ 
\end{IEEEproof}

\section{\label{sec:Simulation}Simulation}

To validate the proposed switched controller, we performed a simulation
with 5 dynamic agents and 3 obstacles. The parameters used in the
simulation are given by $R_{S}=20$, $\delta_{1}=8,$ $\delta_{2}=2,$
$k=1,$ $\Gamma=10,$ $c_{12}=[0,\,5]^{T},$ $c_{23}=[-5,\,5]^{T}$,
$c_{34}=[-5,\,-5]^{T},$ $c_{45}=[0,\,-5]^{T}.$ Initially the agents
are located within the sensing region of their formation neighbors.
Fig. \ref{fig:1} illustrates that the agents avoid collisions with
other agents and stationary obstacles. Moreover, they eventually achieve
an approximation of their goal formation under arbitrary switching
sequence that satisfies (\ref{eq:graph condition}).

\begin{center}
\begin{figure}[H]
\begin{centering}
\includegraphics[scale=0.45]{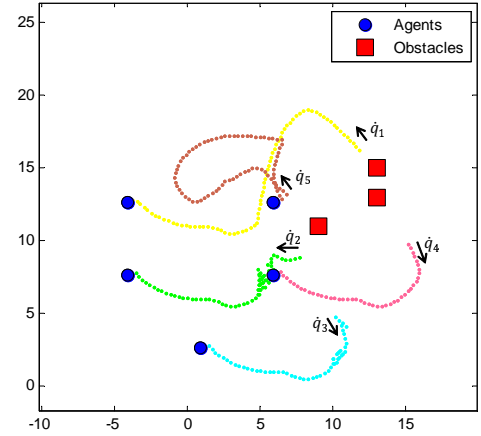}
\par\end{centering}

\caption{\label{fig:1}Trajectories of dynamic agents achieving formation configuration.}
\end{figure}

\par\end{center}

\begin{figure}[H]
\begin{centering}
\includegraphics[scale=0.35]{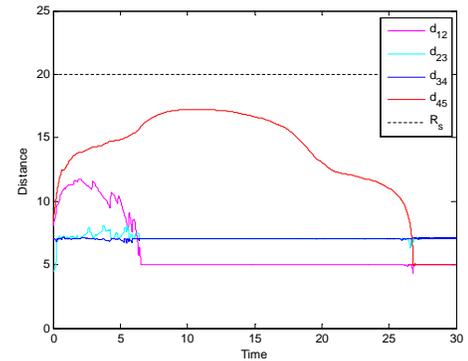}
\par\end{centering}

\caption{\label{fig:2}$d_{ij}$ and $R_{s}$}

\end{figure}

As indicated in Fig. \ref{fig:2}, $d_{ij}$ can increase during operation.
However, these distances always remain smaller than the sensing range
$R_{s}$ (i.e., remain connected). Recall that the relative distance
in our goal formations are given by $\left\Vert c_{12}\right\Vert =\left\Vert c_{45}\right\Vert =5,$
and $\left\Vert c_{23}\right\Vert =\left\Vert c_{34}\right\Vert =5\sqrt{2}.$
Fig. \ref{fig:2} indicates that the final distances approximate these
values, and the position errors remain sufficiently small.

\section{Conclusion}

A switched controller is developed to achieve convergence of a network
formation using only local feedback under both limited and intermittent
sensing. At the same time, network connectivity is maintained and
collisions between agents and obstacles are avoided. A common Lyapunov
function approach is used to ensure convergence under an arbitrary
switching sequence. Moreover the entire formation configuration converges
globally, if the switching signal satisfies (\ref{eq:graph condition}).
The neighborhood of convergence can be made arbitrarilly small with
sufficiently large gains. Finally, the effectiveness of the proposed
controller is verified by simulation results.

\appendix{}

This section develops a sufficient condition for $\Gamma\left(\nabla_{q_{i}}\varphi_{i}\right)^{T}\left(\sum_{j=1}^{N}\nabla_{q_{i}}\varphi_{j}\right)>0,$
so that $\dot{V}$ in (\ref{eq:V_dot3}) is negative definite almost
everywhere. We consider the equation $\left(\nabla_{q_{i}}\varphi_{i}\right)^{T}\left(\sum_{j=1}^{N}\nabla_{q_{i}}\varphi_{j}\right)=\left(\frac{\beta_{i}\left(\nabla_{q_{i}}\gamma_{i}\right)-\frac{\gamma_{i}}{k}\left(\nabla_{q_{i}}\beta_{i}\right)}{\left(\gamma_{i}^{k}+\beta_{i}\right)^{\frac{1}{k}+1}}\right)^{T}\left(\sum_{j=1}^{N}\frac{\beta_{j}\left(\nabla_{q_{i}}\gamma_{j}\right)-\frac{\gamma_{j}}{k}\left(\nabla_{q_{i}}\beta_{j}\right)}{\left(\gamma_{j}^{k}+\beta_{j}\right)^{\frac{1}{k}+1}}\right),$
and decompose this into smaller pieces. Using \cite{Kan.Dani.ea2012}
as inspiration, it is sufficient to ensure the term 
\begin{equation}
A^{T}C-\frac{\left(\left\Vert B\right\Vert \left\Vert C\right\Vert +\left\Vert A\right\Vert \left\Vert D\right\Vert \right)}{k}-\frac{\left\Vert B\right\Vert \left\Vert D\right\Vert }{k^{2}}>0,\label{eq:ABCD ineq}
\end{equation}
 where $A,\, B,\, C,\, D\in\mathbb{R}^{2}$ are from the numerator
terms of $\left(\nabla_{q_{i}}\varphi_{i}\right)^{T}\left(\sum_{j=1}^{N}\nabla_{q_{i}}\varphi_{j}\right)$
and are defined as $A\triangleq\beta_{i}\left(\nabla_{q_{i}}\gamma_{i}\right),$
$B\triangleq\gamma_{i}\left(\nabla_{q_{i}}\beta_{i}\right),$ $C\triangleq\sum_{j=1}^{N}\beta_{j}\left(\nabla_{q_{i}}\gamma_{j}\right),$
and $D\triangleq\sum_{j=1}^{N}\gamma_{j}\left(\nabla_{q_{i}}\beta_{j}\right).$
We now proceed to find upper bounds for $\left\Vert A\right\Vert ^{2},$
$\left\Vert B\right\Vert ^{2},$ $\left\Vert C\right\Vert ^{2},$
and $\left\Vert D\right\Vert ^{2}$ so that we can satisfy $A^{T}C-\frac{\left\Vert B\right\Vert ^{2}+\left\Vert C\right\Vert ^{2}+\left\Vert A\right\Vert ^{2}+\left\Vert D\right\Vert ^{2}}{2k}-\frac{\left\Vert B\right\Vert ^{2}+\left\Vert D\right\Vert ^{2}}{2k^{2}}>0,$
which is the upper bound of (\ref{eq:ABCD ineq}). 
\begin{property}
\label{Property1}$\left\Vert A\right\Vert ^{2}\leq4\beta_{i}^{2}\left|\mathcal{N}_{i}^{f}\right|\gamma_{i}.$\end{property}
\begin{IEEEproof}
By definition $A=\beta_{i}\left(\nabla_{q_{i}}\gamma_{i}\right)=\beta_{i}\left(2\underset{j\in\mathcal{N}_{i}^{f}}{\sum}\left(q_{i}-q_{j}-c_{ij}\right)\right)=2\beta_{i}\underset{j\in\mathcal{N}_{i}^{f}}{\sum}\left(q_{i}-q_{j}-c_{ij}\right),$
from which it follows that $\left\Vert A\right\Vert ^{2}=$ 
\begin{align}
\left\Vert A\right\Vert ^{2} & =4\beta_{i}^{2}\left\Vert \underset{j\in\mathcal{N}_{i}^{f}}{\sum}\left(q_{i}-q_{j}-c_{ij}\right)\right\Vert ^{2}.\label{eq:A2}
\end{align}
 Taking $X_{j}\triangleq[x_{j1},\, x_{j2}]^{T}$ we can bound the
sum, first by using the triangle inequality as 
\begin{align}
\left\Vert \underset{j\in\mathcal{N}_{i}^{f}}{\sum}X_{j}\right\Vert ^{2} & \leq\left(\underset{j\in\mathcal{N}_{i}^{f}}{\sum}\left\Vert X_{j}\right\Vert \right)^{2}=\left(\underset{j\in\mathcal{N}_{i}^{f}}{\sum}\left\Vert X_{j}\right\Vert \cdot1\right)^{2}.\label{eq:note 1-1}
\end{align}
Next we apply the Cauchy\textendash{}Schwarz inequality, and bound
(\ref{eq:note 1-1}) as 
\begin{align}
\left\Vert \underset{j\in\mathcal{N}_{i}^{f}}{\sum}X_{j}\right\Vert ^{2} & \leq\left(\underset{j\in\mathcal{N}_{i}^{f}}{\sum}\left\Vert X_{j}\right\Vert ^{2}\right)\left(\underset{j\in\mathcal{N}_{i}^{f}}{\sum}1\right)\nonumber \\
 & \leq\left|\mathcal{N}_{i}^{f}\right|\underset{j\in\mathcal{N}_{i}^{f}}{\sum}\left\Vert X_{j}\right\Vert ^{2}.\label{eq:note 3-1}
\end{align}
We can bound $\left\Vert A\right\Vert ^{2}$ by using (\ref{eq:note 3-1})
to establish Property 1. \end{IEEEproof}
\begin{property}
\label{Property2}$\left\Vert B\right\Vert \leq\gamma_{i}\left(\left|\mathcal{N}_{i}^{f}\right|\frac{2}{\delta_{2}}+\left|\mathcal{N}_{i}\cup\mathcal{M}_{i}\right|\frac{2}{\delta_{1}}\right).$\end{property}
\begin{IEEEproof}
Given the definition: $B=\gamma_{i}\left(\nabla_{q_{i}}\beta_{i}\right)=\gamma_{i}\left(\underset{j\in\mathcal{N}_{i}^{f}}{\sum}\left(\nabla_{q_{i}}b_{ij}\right)\overline{b}_{ij}+\underset{k\in\mathcal{N}_{i}\cup\mathcal{M}_{i}}{\sum}\left(\nabla_{q_{i}}B_{ik}\right)\bar{B}_{ik}\right),$
where we take $\bar{B}_{ik}\triangleq\underset{j\in\mathcal{N}_{i}^{f}}{\prod}b_{ij}\underset{h\in\mathcal{N}_{i}\cup\mathcal{M}_{i},\, h\neq k}{\prod}B_{ih},$
since $b_{ij}\,\mbox{and}\, B_{ik}\in\left[0,\,1\right],$ then $\overline{b}_{ij},\,\bar{B}_{ik}\in\left[0,\,1\right].$
Thus, we can develop the following inequality for $\left\Vert B\right\Vert $:
\begin{equation}
\left\Vert B\right\Vert \leq\gamma_{i}\left(\underset{j\in\mathcal{N}_{i}^{f}}{\sum}\left\Vert \nabla_{q_{i}}b_{ij}\right\Vert +\underset{k\in\mathcal{N}_{i}\cup\mathcal{M}_{i}}{\sum}\left\Vert \nabla_{q_{i}}B_{ik}\right\Vert \right).\label{eq:note 4-1}
\end{equation}
 By using (\ref{eq:gradient b_ij}), $\left\Vert \nabla_{q_{i}}b_{ij}\right\Vert \leq\frac{2}{\delta_{2}}.$
In a similar manner,$\left\Vert \nabla_{q_{i}}B_{ik}\right\Vert \leq\frac{2}{\delta_{1}}.$
Property 2 is proven by applying these inequalities term by term to
(\ref{eq:note 4-1}). \end{IEEEproof}
\begin{property}
\label{Property3}$\left\Vert C\right\Vert ^{2}\leq4\left|\mathcal{N}_{i}^{f}\right|\gamma_{i}.$\end{property}
\begin{IEEEproof}
Recall that C is defined as $C\triangleq\sum_{j=1}^{N}\beta_{j}\left(\nabla_{q_{i}}\gamma_{j}\right)=\underset{j\in\mathcal{V}}{\sum}\beta_{j}\left(\nabla_{q_{i}}\gamma_{j}\right)=\underset{j\in\mathcal{N}_{i}^{f}}{\sum}\beta_{j}\left(\nabla_{q_{i}}\gamma_{j}\right)+\underset{j\in\mathcal{V}\setminus\mathcal{N}_{i}^{f}}{\sum}\beta_{j}\left(\nabla_{q_{i}}\gamma_{j}\right).$
Since the graph is undirected, whenever $j$ in $\mathcal{N}_{i}^{f},$
we have $i$ in $\mathcal{N}_{j}^{f}.$ Therefore, for any agent $i$
in $\mathcal{N}_{j}^{f}$ 
\begin{align}
\nabla_{q_{i}}\gamma_{j} & =\nabla_{q_{i}}\left(\left\Vert q_{j}-q_{i}-c_{ji}\right\Vert ^{2}\right)\nonumber \\
 & \qquad\qquad\;+\nabla_{q_{i}}\left(\underset{h\neq i}{\underset{h\in\mathcal{N}_{j}^{f}}{\sum}}\left\Vert q_{j}-q_{h}-c_{jh}\right\Vert ^{2}\right)\nonumber \\
 & =\nabla_{q_{i}}\left(\underset{i\in\mathcal{N}_{j}^{f}}{\sum}\left\Vert q_{j}-q_{i}-c_{ji}\right\Vert ^{2}\right)\nonumber \\
 & =-2\left(q_{j}-q_{i}-c_{ji}\right)=2\left(q_{i}-q_{j}-c_{ij}\right).\label{eq:property C bound}
\end{align}
 By using (\ref{eq:property C bound}) 
\begin{align}
\underset{j\in\mathcal{N}_{i}^{f}}{\sum}\beta_{j}\left(\nabla_{q_{i}}\gamma_{j}\right) & =\underset{j\in\mathcal{N}_{i}^{f}}{\sum}\beta_{j}\left(2\left(q_{i}-q_{j}-c_{ij}\right)\right)\nonumber \\
 & =2\underset{j\in\mathcal{N}_{i}^{f}}{\sum}\beta_{j}\left(q_{i}-q_{j}-c_{ij}\right).\label{eq:note5-1}
\end{align}

On the contrary, if $j$ is not in $\mathcal{N}_{i}^{f},$ then $\nabla_{q_{i}}\gamma_{j}=\nabla_{q_{i}}\left(\underset{i\in\mathcal{N}_{j}^{f}}{\sum}\left\Vert q_{j}-q_{i}-c_{ji}\right\Vert ^{2}\right)=0,$
which indicates that $\underset{j\in\mathcal{V}\setminus\mathcal{N}_{i}^{f}}{\sum}\beta_{j}\left(\nabla_{q_{i}}\gamma_{j}\right)=0.$
Finally, using (\ref{eq:note5-1}) 
\[
C=\underset{j\in\mathcal{N}_{i}^{f}}{\sum}\beta_{j}\left(\nabla_{q_{i}}\gamma_{j}\right)=\underset{j\in\mathcal{N}_{i}^{f}}{\sum}\beta_{j}\left(2\left(q_{i}-q_{j}-c_{ij}\right)\right).
\]
According to $\beta_{j}\in\left[0,\,1\right],\,\forall j\in\mathcal{V},$
$\left\Vert C\right\Vert $ can be bounded by $\left\Vert C\right\Vert \leq2\left\Vert \underset{j\in\mathcal{N}_{i}^{f}}{\sum}\left(q_{i}-q_{j}-c_{ij}\right)\right\Vert ,$
and $\left\Vert C\right\Vert ^{2}$ can be further bounded by 
\begin{align*}
\left\Vert C\right\Vert ^{2} & \leq4\left\Vert \underset{j\in\mathcal{N}_{i}^{f}}{\sum}\left(q_{i}-q_{j}-c_{ij}\right)\right\Vert ^{2}.
\end{align*}
By using (\ref{eq:note 3-1}), $\left\Vert C\right\Vert ^{2}$ can
be bounded by 
\[
\left\Vert C\right\Vert ^{2}\leq4\left|\mathcal{N}_{i}^{f}\right|\underset{j\in\mathcal{N}_{i}^{f}}{\sum}\left\Vert \left(q_{i}-q_{j}-c_{ij}\right)\right\Vert ^{2}=4\left|\mathcal{N}_{i}^{f}\right|\gamma_{i}.
\]
\end{IEEEproof}
\begin{property}
\label{Property4}$\left\Vert D\right\Vert \leq\left(\frac{2}{\delta_{2}}+\frac{2}{\delta_{1}}\right)\sum_{j=1}^{N}\gamma_{j}.$\end{property}
\begin{IEEEproof}
By using the definition of $D=\sum_{j=1}^{N}\gamma_{j}\left(\nabla_{q_{i}}\beta_{j}\right)$
and applying the same inequalities used in the proof of Property 2
\begin{align*}
\left\Vert D\right\Vert  & =\left\Vert \sum_{j=1}^{N}\gamma_{j}\left(\nabla_{q_{i}}\beta_{j}\right)\right\Vert \leq\sum_{j=1}^{N}\left\Vert \gamma_{j}\right\Vert \left\Vert \left(\nabla_{q_{i}}\beta_{j}\right)\right\Vert \\
 & \leq\sum_{j=1}^{N}\left\Vert \gamma_{j}\right\Vert \left(\frac{2}{\delta_{2}}+\frac{2}{\delta_{1}}\right).
\end{align*}
 Since $\gamma_{j}\in\mathbb{R}_{\geq0}$ (i.e., $\gamma_{j}=\left\Vert \gamma_{j}\right\Vert $),
$\left\Vert D\right\Vert $ can be further bounded by $\left\Vert D\right\Vert \leq\left(\frac{2}{\delta_{2}}+\frac{2}{\delta_{1}}\right)\sum_{j=1}^{N}\gamma_{j}.$\end{IEEEproof}
\begin{property}
\label{Property5}$\gamma_{i}\leq\left|\mathcal{N}_{i}^{f}\right|\left(R_{s}+\bar{c}_{i}\right)^{2},$
where $\bar{c}_{i}=\underset{j\in\mathcal{N}_{i}^{f}}{\max}\left\Vert c_{ij}\right\Vert .$\end{property}
\begin{IEEEproof}
From (\ref{eq:connectivity eq}), $\left\Vert q_{i}-q_{j}\right\Vert \leq R_{s},\, j\in\mathcal{N}_{i}^{f},$
then $\left\Vert q_{i}-q_{j}-c_{ij}\right\Vert \leq\left\Vert q_{i}-q_{j}\right\Vert +\left\Vert c_{ij}\right\Vert \leq R_{s}+\left\Vert c_{ij}\right\Vert ,$
which implies $\gamma_{i}=\underset{j\in\mathcal{N}_{i}^{f}}{\sum}\left\Vert q_{i}-q_{j}-c_{ij}\right\Vert ^{2}\leq\underset{j\in\mathcal{N}_{i}^{f}}{\sum}\left\Vert R_{s}+\left\Vert c_{ij}\right\Vert \right\Vert ^{2}.$
By choosing the $\bar{c}_{i}=\underset{j\in\mathcal{N}_{i}^{f}}{\max}\left\Vert c_{ij}\right\Vert ,$
then 
\[
\gamma_{i}\leq\left|\mathcal{N}_{i}^{f}\right|\left(R_{s}+\bar{c}_{i}\right)^{2}.
\]

\end{IEEEproof}
Recall that our goal is to establish (\ref{eq:ABCD ineq}). We will
instead establish this for the smaller equation obtained by way of
Young's inequality: $A^{T}C-\frac{\left\Vert B\right\Vert ^{2}+\left\Vert C\right\Vert ^{2}+\left\Vert A\right\Vert ^{2}+\left\Vert D\right\Vert ^{2}}{2k}-\frac{\left\Vert B\right\Vert ^{2}+\left\Vert D\right\Vert ^{2}}{2k^{2}}\leq A^{T}C-\frac{\left(\left\Vert B\right\Vert \left\Vert C\right\Vert +\left\Vert A\right\Vert \left\Vert D\right\Vert \right)}{k}-\frac{\left\Vert B\right\Vert \left\Vert D\right\Vert }{k^{2}}.$
By using the upper bounds established in Property 1-4, we find: 
\begin{align}
 & A^{T}C-\frac{\left\Vert B\right\Vert ^{2}+\left\Vert C\right\Vert ^{2}+\left\Vert A\right\Vert ^{2}+\left\Vert D\right\Vert ^{2}}{2k}-\frac{\left\Vert B\right\Vert ^{2}+\left\Vert D\right\Vert ^{2}}{2k^{2}}\nonumber \\
 & \geq4\underline{\beta}\left\Vert \underset{j\in\mathcal{N}_{i}^{f}}{\sum}\left(q_{i}-q_{j}-c_{ij}\right)\right\Vert ^{2}-\frac{\rho_{1,i}}{2k}-\frac{\rho_{2,i}}{2k^{2}},\label{eq:v_dot in preliminary}
\end{align}
where $\rho_{1,i}$ and $\rho_{2,i}$ are defined below (\ref{eq:V_dot4-1}).
In other words, if the right hand side of (\ref{eq:v_dot in preliminary})
is positive, then $A^{T}C-\frac{\left(\left\Vert B\right\Vert \left\Vert C\right\Vert +\left\Vert A\right\Vert \left\Vert D\right\Vert \right)}{k}-\frac{\left\Vert B\right\Vert \left\Vert D\right\Vert }{k^{2}}>0.$
In addition, we would have a sufficient condition for $\left(\nabla_{q_{i}}\varphi_{i}\right)^{T}\left(\sum_{j=1}^{N}\nabla_{q_{i}}\varphi_{j}\right)>0.$
Thus by (\ref{eq:v_dot in preliminary}) it suffices to show 
\begin{align}
 & \left(4\underline{\beta}\left\Vert \underset{j\in\mathcal{N}_{i}^{f}}{\sum}\left(q_{i}-q_{j}-c_{ij}\right)\right\Vert ^{2}-\frac{\rho_{1,i}}{2k}-\frac{\rho_{2,i}}{2k^{2}}\right)>0.\label{eq:conclusion-preliminary}
\end{align}
 Based on Property \ref{Property5}, $\gamma_{i}$ can be bounded
above by a constant, which means $\rho_{1,i}$ and $\rho_{2,i}$ both
have upper bounds of $\overline{\rho}_{1}$ and $\overline{\rho}_{2}$
defined below (\ref{eq:Gamma UUB}). In addition, in (\ref{eq:conclusion-preliminary})
$\underline{\beta}\in\mathbb{R}$ is a positive constant defined below
(\ref{eq:Gamma UUB}).

\bibliographystyle{IEEEtran}
\bibliography{C:/Users/Tenghu/Desktop/bibtex/bib/ncrbibs/encr,C:/Users/Tenghu/Desktop/bibtex/bib/ncrbibs/master,C:/Users/Tenghu/Desktop/bibtex/bib/ncrbibs/ncr}

\end{document}